\documentclass[11pt]{article}

\usepackage[T1]{fontenc}
\usepackage[utf8]{inputenc}
\usepackage{lmodern}
\usepackage{amsmath,amssymb,mathtools,amsthm}
\usepackage{geometry}
\usepackage{microtype}
\usepackage{booktabs}
\usepackage{authblk}
\usepackage{hyperref}

\hypersetup{hidelinks}
\numberwithin{equation}{section}

\newtheorem{definition}{Definition}
\newtheorem{proposition}{Proposition}

\newcommand{\F}{\mathbb{F}}
\newcommand{\N}{\mathbb{N}}
\newcommand{\Part}{\operatorname{Part}}
\newcommand{\CommonRef}{\operatorname{Ref}}

\title{Prestatability in the Macroscopic Limit}
\author{Andrea Palessandro\thanks{andrea.palessandro@gmail.com}}
\affil{\small Societ\`a Italiana di Fisica}
\date{}

\begin{document}

\maketitle

\begin{abstract}                                                                                                                                                                                                                                                                                                                                                                                                                    
    \noindent 
    We examine whether the variables required for a future macroscopic description of a given physical system can themselves be stated in advance. 
    Representing macroscopic descriptions as coarse-grainings of a fixed microscopic phase space, we distinguish \emph{formal prestatability}, 
    concerning the existence of an effective complete vocabulary of admissible macroscopic distinctions, from \emph{dynamical prestatability}, 
    concerning whether the future relevance of those distinctions is decidable. Both obstructions disappear for finite microscopic state spaces. 
    In the infinite limit, exhaustive formal prestatability can fail, while a restricted predictive architecture may remain formally pre-statable. 
    We construct such an architecture and show that, under universal dynamics, the future relevance of a candidate variable is equivalent to the halting problem.
    Thus a fixed microscopic phase space can support macroscopic descriptions whose future variables are not fully pre-statable.
\end{abstract}

\section{Introduction}

A central question in the study of complex systems is not merely whether the future
values of known variables can be predicted, but whether the variables needed
to describe the future can themselves be stated in advance.

This distinction is central to the account of biological evolution developed
by Kauffman and collaborators
\cite{Kauffman1993,Kauffman2000,LongoMontevilKauffman2012,
    LongoMontevil2013,KauffmanRoli2021,
    Kauffman2022,KauffmanRoli2023,RoliSucciKauffman2026}. In the usual
mathematical description of a physical system, the relevant variables and
their possible values are specified before the dynamics is considered. They
define a phase space, and the subsequent evolution is represented as a
trajectory within that space. The trajectory may be extremely difficult to
predict, chaotic, stochastic, or computationally irreducible, but the space
of possible states is nevertheless given in advance.

Kauffman and collaborators argue that biological evolution presents a
different situation. Evolution goes beyond generating unexpected values of
pre-existing variables: it gives rise to new phenotypes, functions, niches,
and affordances that can themselves become relevant variables for the subsequent
evolution of the biosphere
\cite{TriaEtAl2014,TaylorEtAl2016,PackardEtAl2019}. In their account, these possibilities cannot be
exhaustively specified beforehand. In particular, an affordance is not an
intrinsic property of an object considered in isolation, but a possible use
arising from the relation between an organism, its environment, and its
possible actions \cite{Gibson1979}. As organisms and environments change, new such relations
can arise. Selection may therefore reveal, only after the fact, features of
organisms and environments that become relevant to a newly constituted
niche.

This motivates Kauffman's distinction between \emph{unpredictability} and
\emph{unprestatability}. An unpredictable system may evolve within a phase
space that is completely specified in advance: what is unknown is which
trajectory will occur. An unprestatable system, by contrast, is one for which
the relevant space of possibilities cannot itself be completely specified
before the evolution unfolds. In this stronger sense, the problem is not only
that we cannot predict what will happen, but that we cannot prestate all of
the variables in terms of which future possibilities will eventually be
described.

The present paper examines this claim by separating two notions that are
usually left implicit: \emph{formal prestatability} and \emph{dynamical
    prestatability}.

Before examining these notions in detail, we first want to clarify the meaning of a macroscopic description. 
At the microscopic level, a single phase space is in principle sufficient to
represent the complete state of the system. At the macroscopic level, however,
one usually works with reduced descriptions that retain only those distinctions
that matter for a particular scale, task, or regime. If one assumes that the microscopic phase space $\Omega$ is fixed, a
macroscopic description may be represented by a coarse-graining of $\Omega$,
that is, by a partition of microscopic states into classes that are
indistinguishable at the macroscopic level
\cite{Atmanspacher2016,ShaliziMoore2025}. A collection of macroscopic
variables provides one way of generating such a partition: two microscopic
states belong to the same block whenever all of the variables take the same
values on them.

Different macroscopic questions may therefore require different coarse-grainings of the same
underlying microscopic state space. A description appropriate for predicting
one observable, identifying one function, or characterizing one regime need not
remain appropriate when the system enters another context. In this sense, a
macroscopic phase space is not merely a smaller copy of the microscopic one,
but a task- and context-dependent representation of it. Once macroscopic description is understood in this way, the natural question
is no longer only which state the system will occupy, but also which
coarse-graining of the microscopic space will be required to describe it.

This formulation reveals two distinct ways in which the future macroscopic
description may fail to be pre-statable.\footnote{Throughout, \emph{pre-statable}
    means effectively specifiable by a finite procedure in the sense made
    precise below, not merely mathematically definable or nameable.} The first concerns the space of
possible descriptions itself. Given a class of macroscopic coarse-grainings
regarded as admissible, can one specify in advance an effective vocabulary of
elementary distinctions from which every admissible description can be
constructed? We call this question one of \emph{formal prestatability}. A
failure of formal prestatability means that no effectively specifiable
vocabulary is sufficient to exhaust the macroscopic descriptions allowed by
the theory.

The second question arises even if such a vocabulary has already been given.
Suppose that every macroscopic variable that could possibly be required
belongs to an explicitly specified family. Can one nevertheless determine,
for an arbitrary member of that family, whether it will ever become relevant
as the system evolves? We call this \emph{dynamical prestatability}. 
It differs from ordinary unpredictability in the object being determined. 
Ordinary prediction takes a descriptive framework as fixed and asks which states 
or values will occur within it. 
Dynamical prestatability asks instead whether the future descriptive framework itself (the set of variables that will have to be retained) 
can be stated in advance.
Its failure therefore indicates that the future macroscopic description cannot be fully determined from the dynamics ahead of time,
even when the complete set of possible variables is known.

Nevertheless, neither obstruction can occur in the strict sense developed here for a genuinely finite,
effectively specified deterministic system. If the microscopic state space is
finite, there are only finitely many partitions of it, so every possible
coarse-graining can in principle be enumerated. If the dynamical state space
is also finite, every trajectory eventually repeats, so the question of
whether a given variable will ever become relevant can likewise be settled by
a finite exhaustive search. Computational complexity may make such procedures
hopeless in practice, so finite systems may still be operationally
unprestatable under physically available computational resources, but no strict
mathematical obstruction to prestatability arises.
Strict unprestatability therefore enters only when one passes to an infinite
or effectively unbounded system. 

The first obstruction appears already at the level of possible
coarse-grainings. For an infinite microscopic phase space, the collection of
all partitions of $\Omega$ does not admit an effective exhaustive
enumeration. If every mathematically possible coarse-graining were taken to
represent an admissible macroscopic description, no effective vocabulary
could therefore be complete. We refer to this as a failure of
\emph{exhaustive formal prestatability}. This is a genuine obstruction, but
it must be interpreted with care: arbitrary partitions of a microscopic
state space do not in general correspond to physically meaningful variables.
Consequently, the non-enumerability of the full partition space does not by
itself establish that the macroscopic descriptions relevant to a particular
physical problem are unprestatable.

This motivates the second part of the analysis. We restrict attention to a
specific operational class of macroscopic descriptions, defined by a fixed
prediction problem, and construct an explicit effective vocabulary that is
complete for that class. Formal prestatability is therefore granted from the
outset. We then ask whether the future subset of variables that become
relevant can be determined from the dynamics.
Two models are constructed with the same infinite microscopic state space,
the same observable, the same candidate macroscopic variables, and the same
criterion of predictive relevance. They differ only in the dynamics that
determines which microscopic distinctions are brought into the macroscopic
description. In the first model, the relevant variables change indefinitely,
but the entire future pattern can be stated in advance. The evolving
macroscopic phase space is therefore pre-statable. 
In the second model, the dynamics performs a universal computation. Whether
a candidate variable ever becomes relevant is then equivalent to whether a
corresponding computation eventually halts. The complete vocabulary of
possible variables is explicitly known, and the macroscopic description at
every specified finite time is computable, yet there exists no algorithm that
decides, for every candidate variable, whether it will ever be needed. The
system is formally pre-statable but dynamically unprestatable.

\section{Macroscopic Descriptions and Formal Prestatability}
\label{sec:formal}

We introduce the mathematical framework used throughout the paper. The
basic idea is to represent a macroscopic description as a coarse-graining of the microscopic state space.

Let $\Omega$ denote the microscopic state space of a system. A point $\omega\in\Omega$
specifies a complete microscopic state. At this level, two distinct points of
$\Omega$ represent physically distinct microscopic configurations.

A macroscopic description generally contains less information. It identifies
microscopic states whenever the distinctions between them are irrelevant for
the description under consideration. Mathematically, this is naturally
represented by a partition of $\Omega$.

A partition $P\in\Part(\Omega)$
is a collection of disjoint subsets, or blocks, whose union is $\Omega$.
Microscopic states belonging to the same block are regarded as
indistinguishable at the macroscopic level. Equivalently, $P$ defines an
equivalence relation
\begin{equation}                                                                                                                                                                                                                                                                                                                                                                                                                                                                                                                                                                                                                                                                                                                                                                                                                                                                                                                            
    \omega\sim_P\omega'
\end{equation}
whenever $\omega$ and $\omega'$ lie in the same block of $P$.
The corresponding macroscopic state space is the quotient
\begin{equation}                                                                                                                                                                                                                                                                                                                                                                                                                                                                                                                                                                                                                                                                                                                                                                                                                                                                                                                            
    \mathcal M_P
    =
    \Omega/{\sim_P}.
    \label{eq:macro-quotient}
\end{equation}
A point of $\mathcal M_P$ is a class of microscopic states that the chosen macroscopic description
does not distinguish
\cite{Atmanspacher2016,ShaliziMoore2025}.

A macroscopic variable provides a particularly simple way of generating a
partition. Let                                                                                                                                                                                                                                                                                                                                                                                                                                                                                                                                                                                                                                                              
$X:\Omega\longrightarrow\mathcal X$
be a variable defined on microscopic states. It induces the partition
\begin{equation}                                                                                                                                                                                                                                                                                                                                                                                                                                                                                                                                                                                                                                                                                                                                                                                                                                                                                                                            
    P_X
    =
    \left\{
    X^{-1}(x):
    x\in\mathcal X,\;
    X^{-1}(x)\neq\varnothing
    \right\}.
    \label{eq:variable-partition}
\end{equation}
Two microscopic states belong to the same block of $P_X$ precisely when they
give the same value of $X$:
\begin{equation}                                                                                                                                                                                                                                                                                                                                                                                                                                                                                                                                                                                                                                                                                                                                                                                                                                                                                                                            
    \omega\sim_{P_X}\omega'
    \quad\Longleftrightarrow\quad
    X(\omega)=X(\omega').
\end{equation}
Thus specifying a macroscopic variable is equivalent, for our purposes, to
specifying which microscopic distinctions that variable is capable of
detecting.

Usually a macroscopic description involves more than one variable. Suppose,
for example, that we retain
$X_1,\ldots,X_k$.
Two microscopic states are then macroscopically indistinguishable only if
they agree on every one of these variables:
\begin{equation}                                                                                                                                                                                                                                                                                                                                                                                                                                                                                                                                                                                                                                                                                                                                                                                                                                                                                                                            
    X_i(\omega)=X_i(\omega')
    \qquad
    \text{for all }i=1,\ldots,k.
\end{equation}
The resulting partition is the common refinement of the partitions generated
by the individual variables.
Given a finite family $\mathcal Q\subseteq\Part(\Omega)$, we define its common
refinement by
\begin{equation}                                                                                                                                                                                                                                                                        
    \CommonRef(\mathcal Q)
    =
    \left\{
    \bigcap_{P\in\mathcal Q} B_P
    :
    B_P\in P\text{ for every }P\in\mathcal Q,
    \quad
    \bigcap_{P\in\mathcal Q} B_P\neq\varnothing
    \right\}.
    \label{eq:common-refinement}
\end{equation}
For the empty family, we set $\CommonRef(\varnothing)=\{\Omega\}$.
The refinement retains every distinction made by any partition in $\mathcal Q$.

It is useful to state the corresponding ordering explicitly. We say that a
partition $P$ \emph{refines} a partition $Q$ if every block of $P$ is
contained in a block of $Q$. In that case, $P$ distinguishes at least as many
microscopic states as $Q$. Thus a finer partition corresponds to a more
informative macroscopic description, while a coarser partition discards more
microscopic information.

A simple example makes the construction transparent. Consider
\begin{equation*}                                                                                                                                                                                                                                                                                                                                                                                                                                                                                                                                                                                                                                                                                                                                                                                                                                                                                                                            
    \Omega
    =
    \F_2^2
    =
    \{(0,0),(0,1),(1,0),(1,1)\}.
\end{equation*}
Let
\[                                                                                                                                                                                                                                                                                                                                                                                                                                                                                                                                                                                                                                                                                                                                                                                                                                                                                                                            
    X_1(\omega_1,\omega_2)=\omega_1,
    \qquad
    X_2(\omega_1,\omega_2)=\omega_2.
\]
The variable $X_1$ separates states according to the first coordinate,
whereas $X_2$ separates them according to the second. Each variable by itself
produces a two-block partition. Their common refinement,
$\CommonRef(\{P_{X_1},P_{X_2}\})$,
distinguishes all four microscopic states.
The same microscopic space can therefore support descriptions of different
resolution. One may retain neither distinction, only the first, only the
second, or both.

The set $\Part(\Omega)$
contains every mathematically possible partition of $\Omega$. There is no
reason, however, to regard every such partition as a physically meaningful
macroscopic description. A particular physical problem will generally select only some of them. We
therefore introduce an \emph{admissible class}
\begin{equation}                                                                                                                                                                                                                                                                                                                                                                                                                                                                                                                                                                                                                                                                                                                                                                                                                                                                                                                            
    \mathcal A
    \subseteq
    \Part(\Omega).
    \label{eq:admissible-class}
\end{equation}
The meaning of admissibility depends on the problem being studied. A
partition might be admissible because it corresponds to experimentally
accessible observables, because it is sufficient for a given prediction or
control task, because it captures a relevant functional distinction, or
because it satisfies some other operational criterion.

This distinction between $\Part(\Omega)$
and $\mathcal A$
will be important. The former is the ambient mathematical space of all
possible coarse-grainings. The latter contains only the coarse-grainings that
the physical description actually allows. In the construction developed in
Section~\ref{sec:model}, admissibility will not be chosen arbitrarily. It will
be defined by a fixed prediction problem: a macroscopic
description will be admissible when it retains exactly the microscopic
distinctions required to determine a specified future observable.

We can now formulate the first notion of prestatability.
Suppose that the elementary distinctions available to the theory are
represented by a family
\begin{equation}                                                                                                                                                                                                                                                                                                                                                                                                                                                                                                                                                                                                                                                                                                                                                                                                                                                                                                                            
    \mathfrak P
    =
    \{P_n\}_{n\in\N}
    \subseteq
    \Part(\Omega).
    \label{eq:general-vocabulary}
\end{equation}
We refer to $\mathfrak P$ as a \emph{macroscopic vocabulary}. Its members
should be thought of as elementary variables or distinctions that may be
combined to form more detailed macroscopic descriptions.

From such a vocabulary we can construct all partitions obtained by retaining
finitely many of its elements. We denote this finite-refinement closure by
\begin{equation}                                                                                                                                                                                                                                                                                                                                                                                                                                                                                                                                                                                                                                                                                                                                                                                                                                                                                                                            
    \langle\mathfrak P\rangle_{\mathrm{fin}}
    =
    \left\{
    \CommonRef(\mathcal Q)
    :
    \mathcal Q\subseteq\mathfrak P,
    \ |\mathcal Q|<\infty
    \right\}.
    \label{eq:refinement-closure}
\end{equation}
Thus $Q\in\langle\mathfrak P\rangle_{\mathrm{fin}}$
means that the macroscopic description $Q$ can be expressed using finitely
many distinctions drawn from the vocabulary.\footnote{The restriction to finite refinements is deliberate. The vocabulary itself may be infinite, but any particular macroscopic description is required to involve only finitely many elementary distinctions. This ensures that an admissible description is itself finitely specifiable and prevents an infinite amount of descriptive information from being hidden inside a single macroscopic state space.}
For prestatability, however, merely asserting the existence of such a family
is not enough. The vocabulary itself must be specifiable.
\begin{definition}[Effective vocabulary]
    \label{def:effective-vocabulary}
    A family
    \[                                                                                                                                                                                                                                                                                                                                                                                                                                                                                                                                                                                                                                                                                                                                                                                                                                                                                                                            
        \mathfrak P=\{P_n\}_{n\in\N}
    \]
    is an \emph{effective vocabulary} if there exists a uniform effective
    procedure which, given $n$, produces a finite specification of $P_n$.
\end{definition}
The precise form of such a specification depends on the model. For a
coordinate variable, for example, the integer identifying the coordinate is
already a finite specification. More complicated settings may require a
formula, an algorithm, or some other finite rule. What matters is that the
members of the vocabulary are not merely asserted to exist: they can be
systematically named and described.

Effectivity alone is still insufficient. An effectively enumerable
vocabulary may omit distinctions that are required by admissible macroscopic
descriptions. We therefore also require completeness.

\begin{definition}[Completeness]
    \label{def:completeness}
    An effective vocabulary $\mathfrak P$ is \emph{complete} for an
    admissible class $\mathcal A$ if
    \begin{equation}                                                                                                                                                                                                                                                                                                                                                                                                                                                                                                                                                                                                                                                                                                                                                                                                                                                                                                                            
        \mathcal A
        \subseteq
        \langle\mathfrak P\rangle_{\mathrm{fin}}.
        \label{eq:vocabulary-completeness}
    \end{equation}
    In other words, every admissible macroscopic description can be
    constructed from finitely many distinctions belonging to the
    pre-specified vocabulary.
\end{definition}

We may now state the first notion of prestatability precisely.

\begin{definition}[Formal prestatability]
    \label{def:formal-prestatability}
    An admissible macroscopic class $\mathcal A$ is
    \emph{formally pre-statable} if it possesses an effective vocabulary
    that is complete for $\mathcal A$.
\end{definition}

The meaning of this definition is straightforward. Formal prestatability asks
whether the theory possesses, in advance, a sufficiently rich language for
describing every macroscopic state space that the problem may require. It
does not ask which of those descriptions will actually occur. That is a
separate dynamical question.

It is important that the definition is relative to the admissible class
$\mathcal A$. If the physical problem allows only a restricted family of
coarse-grainings, there is no reason to demand a vocabulary capable of
constructing every member of $\Part(\Omega)$. What must be pre-statable are
the descriptions that can actually become relevant within the theory.

There is nevertheless a useful limiting case. If we take $\mathcal A=\Part(\Omega)$,
we are asking whether every mathematically possible coarse-graining of the
microscopic state space can be generated from an effective vocabulary. We
will refer to this stronger requirement as \emph{exhaustive formal
    prestatability}.
The distinction between ordinary and exhaustive formal prestatability will
be central in Section~\ref{sec:finite}. For a finite microscopic state space, $\Part(\Omega)$ is
itself finite, so exhaustive formal prestatability presents no fundamental
difficulty. For an infinite microscopic space, the situation can be very
different. The first obstruction studied in this paper will arise precisely
there.

The second obstruction is of another kind. Even when an effective complete
vocabulary exists, it may still be impossible to determine which members of
that vocabulary will enter the macroscopic description as the system evolves.
To formulate that question, however, we first need an operational criterion
for deciding when a microscopic distinction is \emph{relevant}. We introduce
such a criterion in the next section by defining the macroscopic description
through a prediction task.

\section{Predictive Relevance and Dynamical Prestatability}
\label{sec:predictive}

The previous section treated macroscopic descriptions abstractly as
coarse-grainings of a fixed microscopic state space. We now need a criterion
for deciding which coarse-graining is relevant in a given situation. Rather
than choosing this criterion by hand, we will derive it from a prediction
problem.

The idea is simple. Suppose that we are interested only in predicting a
particular observable over some finite time interval. Two microscopic states
should then be regarded as macroscopically equivalent whenever they lead to
the same predictions for that observable. Microscopic distinctions that can
never affect the prediction may be discarded, while distinctions that can
affect it must be retained. In this way, the prediction task itself determines
the macroscopic description.

Let $\Omega$ be the microscopic state space of the system under consideration, and let
\begin{equation}                                                                                                                                                                                                                                                                                                                                                                                                                                                                                                                                                                                                                                                                                                                                                                                                                    
    h:\Omega\longrightarrow Y
\end{equation}
be the observable we wish to predict. We also introduce an internal context
space $\mathcal C$. The complete microscopic state space is then
\begin{equation}                                                                                                                                                                                                                                                                                                                                                                                                    
    \widehat\Omega=\mathcal C\times\Omega,
    \qquad
    \widehat\omega=(c,\omega).
    \label{eq:complete-state-space}
\end{equation}
The $\Omega$-component of the update is written as
\begin{equation}                                                                                                                                                                                                                                                                                                                                                                                                                                                                                                                                                                                                                                                                                                                                                                                                                    
    G:\mathcal C\times\Omega\longrightarrow\Omega.
    \label{eq:G-general}
\end{equation}
For the moment, we do not specify how the context itself evolves. We first ask
what macroscopic description is required for a given sequence of contexts.

Fix a finite prediction horizon $H\geq1$ and a context word
\begin{equation}                                                                                                                                                                                                                                                                                                                                                                                                                                                                                                                                                                                                                                                                                                                                                                                                                    
    \mathbf c
    =
    (c_0,c_1,\ldots,c_{H-1})
    \in\mathcal C^H.
\end{equation}
Starting from an initial microscopic state $\omega^{(0)}$,
we define
\begin{equation}                                                                                                                                                                                                                                                                                                                                                                                                                                                                                                                                                                                                                                                                                                                                                                                                                    
    \omega^{(k+1)}
    =
    G(c_k,\omega^{(k)}),
    \qquad
    0\leq k<H.
    \label{eq:state-under-context-word}
\end{equation}
The observable values generated over the prediction horizon are therefore
\begin{equation}                                                                                                                                                                                                                                                                                                                                                                                                                                                                                                                                                                                                                                                                                                                                                                                                                    
    \Psi_{H,\mathbf c}(\omega)
    =
    \bigl(
    h(\omega^{(0)}),
    h(\omega^{(1)}),
    \ldots,
    h(\omega^{(H)})
    \bigr).
    \label{eq:prediction-map-word}
\end{equation}
The map $\Psi_{H,\mathbf c}$ contains precisely the information relevant to
the prediction task. If two microscopic states produce the same sequence of
observable values, then the prediction problem has no reason to distinguish
between them. We therefore define
\begin{equation}                                                                                                                                                                                                                                                                                                                                                                                                                                                                                                                                                                                                                                                                                                                                                                                                                    
    \omega\sim_{H,\mathbf c}\omega'
    \quad\Longleftrightarrow\quad
    \Psi_{H,\mathbf c}(\omega)
    =
    \Psi_{H,\mathbf c}(\omega').
    \label{eq:predictive-equivalence}
\end{equation}
The equivalence classes of this relation form a partition of $\Omega$, which
we denote by
\begin{equation}                                                                                                                                                                                                                                                                                                                                                                                                                                                                                                                                                                                                                                                                                                                                                                                                                    
    \Pi_H(\mathbf c)
    =
    \left\{
    [\omega]_{H,\mathbf c}:
    \omega\in\Omega
    \right\}.
    \label{eq:predictive-partition}
\end{equation}
This partition has a direct operational meaning. Two microscopic states lie
in the same block of $\Pi_H(\mathbf c)$ exactly when they are
indistinguishable with respect to the chosen prediction task. The associated
quotient
\begin{equation}                                                                                                                                                                                                                                                                                                                                                                                                                                                                                                                                                                                                                                                                                                                                                                                                                    
    \mathcal M_H(\mathbf c)
    =
    \Omega/{\sim_{H,\mathbf c}}
    \label{eq:predictive-macrospace}
\end{equation}
is therefore the macroscopic phase space sufficient for predicting the
observable over the horizon $H$ under the context word $\mathbf c$.

It is useful to emphasize the minimality of this construction. Suppose that
two microscopic states $\omega$ and $\omega'$ give different values of
$\Psi_{H,\mathbf c}$. Any description that is sufficient to determine the
complete observable sequence must distinguish between them. Conversely, if
they give the same value of $\Psi_{H,\mathbf c}$, distinguishing them provides
no additional information for the prediction task. Thus $\Pi_H(\mathbf c)$
retains exactly the distinctions needed for the chosen prediction problem,
and no others. Related constructions appear in observability theory,
computational mechanics, and predictive-state representations
\cite{Sontag1979,Nijmeijer1982,HermannKrener1977,
    CrutchfieldYoung1989,Crutchfield1994,ShaliziCrutchfield2001,
    LittmanSuttonSingh2001,CostaEtAl2023}.
The broader problem of compressing microscopic information while preserving
relevant or predictive information also underlies the information bottleneck
and optimal causal inference \cite{TishbyEtAl1999,StillEtAl2010}.

Different context words may require different predictive partitions. The
prediction problem therefore determines a class
\begin{equation}                                                                                                                                                                                                                                                                                                                                                                                                                                                                                                                                                                                                                                                                                                                                                                                                                    
    \mathcal A_H
    =
    \left\{
    \Pi_H(\mathbf c):
    \mathbf c\in\mathcal C^H
    \right\}
    \subseteq
    \Part(\Omega).
    \label{eq:predictive-admissible-class}
\end{equation}
This will be the admissible macroscopic class considered from now on.

The notion of formal prestatability introduced in Section~\ref{sec:formal} can
now be applied directly. The predictive architecture is formally
pre-statable if there exists an effective vocabulary
\begin{equation}                                                                                                                                                                                                                                                                                                                                                                                                                                                                                                                                                                                                                                                                                                                                                                                                                    
    \mathfrak P
    =
    \{P_n\}_{n\in\N}
    \label{eq:effective-vocabulary}
\end{equation}
such that every partition in $\mathcal A_H$ can be constructed from finitely
many members of $\mathfrak P$. In that case, all elementary distinctions that
could be required by the prediction problem are available in advance.

Formal prestatability, however, does not tell us which of those distinctions
will actually become relevant as the system evolves. To ask that question we
must now introduce a dynamics for the context.
Let
\begin{equation}                                                                                                                                                                                                                                                                                                                                                                                                                                                                                                                                                                                                                                                                                                                                                                                                                    
    F:\mathcal C\longrightarrow\mathcal C
    \label{eq:context-map}
\end{equation}
be a deterministic evolution of the internal context, and let $c_0$ be its
initial state.
The subsequent contexts are
\begin{equation}                                                                                                                                                                                                                                                                                                                                                                                                                                                                                                                                                                                                                                                                                                                                                                                                                    
    c_t=F^t(c_0).
\end{equation}
Together, $F$ and $G$ define the autonomous microscopic dynamics
\begin{equation}                                                                                                                                                                                                                                                                                                                                                                                                    
    T:\widehat\Omega\longrightarrow\widehat\Omega,
    \qquad
    T(c,\omega)=\bigl(F(c),G(c,\omega)\bigr).
    \label{eq:autonomous-dynamics}
\end{equation}
At time $t$, prediction over the next $H$ steps depends on the context word
\begin{equation}                                                                                                                                                                                                                                                                                                                                                                                                                                                                                                                                                                                                                                                                                                                                                                                                                    
    \mathbf c_t
    =
    (c_t,c_{t+1},\ldots,c_{t+H-1}),
\end{equation}
and therefore requires the predictive partition
\begin{equation}                                                                                                                                                                                                                                                                                                                                                                                                                                                                                                                                                                                                                                                                                                                                                                                                                    
    \Pi_{H,t}^{F}
    =
    \Pi_H(\mathbf c_t).
    \label{eq:time-predictive-partition}
\end{equation}
The dynamics thus generates a sequence
\begin{equation}                                                                                                                                                                                                                                                                                                                                                                                                                                                                                                                                                                                                                                                                                                                                                                                                                    
    \Pi_{H,0}^{F},
    \Pi_{H,1}^{F},
    \Pi_{H,2}^{F},
    \ldots
\end{equation}
of macroscopic descriptions. Equivalently, these are conditional
coarse-grainings of the microscopic fiber $\{c_t\}\times\Omega$, canonically
identified with $\Omega$.

Suppose now that $\mathfrak P$ is an effective vocabulary that is complete for $\mathcal A_H$. We say that
the elementary distinction $P_n$ is \emph{relevant at time $t$} when the
predictive partition retains the distinction represented by $P_n$. In terms
of partitions, this means
\begin{equation}                                                                                                                                                                                                                                                                                                                                                                                                                                                                                                                                                                                                                                                                                                                                                                                                                    
    \Pi_{H,t}^{F}
    \text{ refines }
    P_n.
    \label{eq:relevance-definition}
\end{equation}
The refinement condition has a simple interpretation. If
$\Pi_{H,t}^{F}$ refines $P_n$, then two microscopic states that differ in the
distinction encoded by $P_n$ cannot in general be identified by the
predictive macroscopic description. The variable represented by $P_n$ must
therefore be retained.

We can now ask which members of the vocabulary will ever become relevant.
Define the future relevance set
\begin{equation}                                                                                                                                                                                                                                                                                                                                                                                                                                                                                                                                                                                                                                                                                                                                                                                                                    
    \mathcal R_H^F(c_0;\mathfrak P)
    =
    \left\{
    n\in\N:
    \Pi_{H,t}^{F}
    \text{ refines }P_n
    \text{ for some }t\geq0
    \right\}.
    \label{eq:future-relevance-general}
\end{equation}
This set describes the variables that will enter the macroscopic description
at least once during the future evolution. It is important to distinguish
this question from ordinary prediction. Ordinary prediction assumes that the
descriptive variables are already fixed and asks which values those variables
will take. Here we ask a different question: which variables will have to
belong to the description at all? The latter question certainly depends on the dynamics, but its object is not
a future microscopic or macroscopic state. Its object is the future
\emph{descriptive scheme}: the set of variables (distinctions) that will have to be
retained as the system evolves.

This leads to the second notion of prestatability.

\begin{definition}[Dynamical prestatability]
    \label{def:dynamical-prestatability}
    Suppose that the predictive admissible class $\mathcal A_H$ is formally
    pre-statable by an effective complete vocabulary
    $\mathfrak P$. The future macroscopic description is
    \emph{dynamically pre-statable} relative to $\mathfrak P$ if membership
    in
    \[                                                                                                                                                                                                                                                                                                                                                                                                                                                                                                                                                                                                                                                                                                                                                                                                                    
        \mathcal R_H^F(c_0;\mathfrak P)
    \]
    is decidable.
\end{definition}
Thus dynamical prestatability asks whether, for an arbitrary candidate
variable $P_n$, there exists a terminating procedure that determines whether
$P_n$ will ever be required by the future macroscopic description.

The distinction between the two notions introduced in
Sections~\ref{sec:formal} and~\ref{sec:predictive} can therefore be
summarized as follows. Formal prestatability concerns the space of
descriptions that the theory permits: do we possess an effective vocabulary
rich enough to construct every admissible macroscopic description?
Dynamical prestatability concerns the trajectory through that space: can we
determine which elements of the available vocabulary will actually become
relevant in the future?

Before constructing examples in which either notion fails, we must first
establish an important baseline. For genuinely finite effectively specified
systems, both questions are decidable in principle. The possibility of strict
unprestatability therefore arises only when one passes to an infinite or
effectively unbounded limit.

\section{Finite Systems and the Infinite-System Limit}
\label{sec:finite}

Before constructing examples of unprestatability, it is important to understand
what can and cannot happen in finite systems. The distinction is
conceptually important because both obstructions introduced in
Sections~\ref{sec:formal} and~\ref{sec:predictive} disappear at
finite size.

Suppose first that the microscopic state space $\Omega$ is finite. Then there
are only finitely many possible partitions of $\Omega$. Indeed, if $|\Omega|=N$,
then
\begin{equation}                                                                                                                                                                                                                                                                                                                                                                                                                                                                                                                                                                                                                                                                                                                                            
    |\Part(\Omega)|=B_N,
\end{equation}  
where $B_N$ is the $N$th Bell number. This number may be enormous, but it is
finite. This already settles the question of exhaustive formal prestatability. Since
every macroscopic description is simply one element of the finite set
$\Part(\Omega)$, one may in principle take the complete vocabulary to contain
all of them. No possible coarse-graining lies beyond finite enumeration.

The same observation applies to dynamical prestatability once the dynamical
state space is finite. Suppose that the context space $\mathcal C$ is finite
and that the context evolution $F$ in Eq.~\eqref{eq:context-map} is
deterministic. Starting from some initial context $c_0$, consider the
trajectory
\begin{equation*}                                                                                                                                                                                                                                                                                                                                                                                                                                                                                                                                                                                                                                                                                                                                                                                                                
    c_0,\;
    F(c_0),\;
    F^2(c_0),\;
    \ldots
\end{equation*}
Because $\mathcal C$ is finite, this sequence cannot visit new states
indefinitely. Eventually some context must repeat. Once that happens, the
subsequent evolution repeats periodically.
Consequently, for any fixed finite prediction horizon $H$, only finitely many
different context words
\[                                                                                                                                                                                                                                                                                                                                                                                                                                                                                                                                                                                                                                                                                                                                                                                                                
    (c_t,c_{t+1},\ldots,c_{t+H-1})
\]
can occur along the future trajectory. Each such word determines one
predictive partition $\Pi_{H,t}^{F}$.
After the context trajectory has entered its periodic part, no genuinely new
predictive partition can appear. The complete future set of relevant macroscopic variables can therefore be
determined by a finite search. One follows the context trajectory until a
previously visited dynamical state is encountered, constructs the predictive
partition at each stage, and records which candidate distinctions are
retained. Once the trajectory repeats, the list is complete.
This gives the finite-state baseline.

\begin{proposition}[Finite-state baseline]
    \label{prop:finite-state-baseline}
    Suppose that $\Omega$ and $\mathcal C$ are finite effectively specified
    sets, and that the microscopic update rule $G$, the observable $h$, and
    the context dynamics $F$ are effectively specified. Then, for every
    finite prediction horizon $H$:
    
    \begin{enumerate}                                                                                                                                                                                                                                                                                                                                                                                                                                                                                                                                                                                                                                                                                                                                                                                                                
        \item the predictive admissible class $\mathcal A_H$ is formally
              pre-statable;
              
        \item the future relevance problem is dynamically decidable.
    \end{enumerate}
\end{proposition}
Proposition~\ref{prop:finite-state-baseline} shows that neither formal nor dynamical unprestatability, in
the strict sense used here, is possible in a finite effectively specified
deterministic system. 
However, finite systems with a large number of degrees of freedom might still be unprestatable in practice. 
A proof
of decidability permits arbitrarily large running times and memory, whereas
real physical systems and observers have only finite computational resources.
If the exhaustive procedure requires more time, memory, or energy than can be
made available within the relevant physical regime, the future macroscopic
description cannot in practice be stated in advance. Finite systems may
therefore be operationally unprestatable even though they are not
unprestatable in the strict mathematical sense defined here
\footnote{We plan to elaborate on this in future work.}.

The obstruction we seek is stronger, and requires some form of unboundedness.
Closely related limit-dependent obstructions are known in mathematical
physics: undecidable or uncomputable behavior has been established for
thermodynamic observables, spectral gaps, phase diagrams, and renormalization
group flows \cite{GuEtAl2009,CubittEtAl2015,BauschEtAl2021,WatsonEtAl2022};
see Ref.~\cite{PeralesEceizaEtAl2025} for a review. These results concern
different decision problems from the one studied here, but they likewise
show how a property that is well defined for every finite instance can become
undecidable in an unbounded limit.

We therefore consider a sequence of finite microscopic systems whose size
increases without bound. A particularly simple example is
\begin{equation}                                                                                                                                                                                                                                                                                                                                                                                                                                                                                                                                                                                                                                                                                                                                                                                                                
    \Omega_N
    =
    \F_2^N
    =
    \left\{
    (g_0,g_1,\ldots,g_{N-1})
    \right\}.
    \label{eq:OmegaN}
\end{equation}
Each microscopic state consists of $N$ binary coordinates. For every finite $N$,
\begin{equation}                                                                                                                                                                                                                                                                                                                                                                                                                                                                                                                                                                                                                                                                                                                                                                                                                
    |\Omega_N|
    =
    2^N,
\end{equation}
so $\Omega_N$ is finite and Proposition~\ref{prop:finite-state-baseline} applies.
The natural unbounded limit is
\begin{equation}                                                                                                                                                                                                                                                                                                                                                                                                                                                                                                                                                                                                                                                                                                                                                                                                                
    \Omega_\infty
    =
    \F_2^{\N}
    =
    \left\{
    (g_0,g_1,g_2,\ldots)
    \right\}.
    \label{eq:Omega-infinity}
\end{equation}
A state in $\Omega_\infty$ contains a countably infinite sequence of binary
coordinates. In the same limit, however, the partition space is uncountable.
For every infinite set $\Omega$ of cardinality $\kappa$, the set of its
partitions (equivalently, its equivalence relations) has cardinality
$|\Part(\Omega)|=2^\kappa$.\footnote{For the upper bound, every equivalence
    relation is a subset of $\Omega\times\Omega$, whose cardinality is $\kappa$.
    For the lower bound, fix $\omega_*\in\Omega$; the proper subsets
    $A\subsetneq\Omega\setminus\{\omega_*\}$ yield $2^\kappa$ distinct two-block
    partitions $\{\{\omega_*\}\cup A,\Omega\setminus(\{\omega_*\}\cup A)\}$.}
For the infinite binary space in Eq.~\eqref{eq:Omega-infinity}, we have
\begin{equation}                                                                                                                                                                                                                                                                                                                                                                                                                                                                                                                                                                                                                                                                                                                                                                                                                
    |\Omega_\infty|
    =
    2^{\aleph_0},
\end{equation}
and therefore
\begin{equation}                                                                                                                                                                                                                                                                                                                                                                                                                                                                                                                                                                                                                                                                                                                                                                                                                
    |\Part(\Omega_\infty)|
    =
    2^{2^{\aleph_0}}.
    \label{eq:Part-Omega-infinity}
\end{equation}
The significance of this result becomes clear when compared with the notion of
an effective vocabulary introduced in Section~\ref{sec:formal}. Any effective
vocabulary can contain at most countably many finitely describable elements:
\[                                                                                                                                                                                                                                                                                                                                                                                                                                                                                                                                                                                                                                                                                                                                                                                                                
    \mathfrak P
    =
    \{P_0,P_1,P_2,\ldots\}.
\]
Moreover, because each admissible macroscopic description is constructed from
only finitely many members of the vocabulary, the finite-refinement closure                                                                                                                                                
$\langle\mathfrak P\rangle_{\mathrm{fin}}$
is itself countable. There are only countably many finite subsets of $\N$, and each such subset
selects only one finite common refinement. Hence a countable effective
vocabulary can generate at most countably many macroscopic descriptions.

This immediately implies the first obstruction.

\begin{proposition}[Failure of exhaustive formal prestatability]
    \label{prop:failure-exhaustive-formal}
    There exists no effective vocabulary that is complete for
    \[                                                                                                                                                                                                                                                                                                                                                                                                                                                                                                                                                                                                                                                                                                                                                                                                                
        \Part(\Omega_\infty).
    \]
    Consequently, the full space of macroscopic coarse-grainings of
    $\Omega_\infty$ is not exhaustively formally pre-statable.
\end{proposition}

The reasoning is simply one of cardinality. The finite-refinement closure of
any effective vocabulary is countable, whereas                                                                                                                                                  
$\Part(\Omega_\infty)$
is uncountable. Most partitions of the infinite microscopic state space
therefore cannot be generated from any effective vocabulary in the sense
defined in \ref{def:formal-prestatability}. This is the first mathematically precise sense in which prestatability can
fail in the infinite-system limit. At every finite size, every possible
coarse-graining can in principle be listed. In the infinite limit, no
effective vocabulary can exhaust the full space of possible partitions.

The conclusion must nevertheless be interpreted carefully. The set
$\Part(\Omega_\infty)$ contains every mathematically conceivable partition of
the microscopic state space, including overwhelmingly many partitions for
which no physical interpretation is apparent. The fact that the complete
partition space is not effectively exhaustible therefore does not establish
that the macroscopic descriptions relevant to a particular physical problem
are themselves formally unprestatable.

A physical theory may select a much smaller admissible class                                                                                                                                                        
$\mathcal A
    \subsetneq
    \Part(\Omega_\infty)$,
and that restricted class may possess a perfectly explicit effective complete
vocabulary. This observation is essential because it separates the two obstructions
introduced in Sections~\ref{sec:formal} and~\ref{sec:predictive}. The failure of exhaustive formal
prestatability follows already from the enormous size of the full partition
space. To demonstrate that the second obstruction is genuinely
distinct, we must construct an infinite system for which the physically
admissible macroscopic descriptions possess an explicit complete vocabulary,
and then ask whether the future dynamics selects from that vocabulary in a
decidable way.

That is the purpose of the model introduced in the next section.

\section{A Formally Pre-statable Predictive Architecture}
\label{sec:model}

We now construct a simple class of systems in which the admissible
macroscopic descriptions possess an explicit effective complete vocabulary.
This will allow us to study dynamical prestatability without relying on the
formal obstruction established in Proposition~\ref{prop:failure-exhaustive-formal}.

We use the infinite binary microscopic state space
$\Omega=\Omega_\infty$ introduced in Eq.~\eqref{eq:Omega-infinity}.
Thus a microscopic state consists of a countably infinite sequence of binary
coordinates. We single out the coordinate $g_0$ as an observable readout. The remaining
coordinates
\[                                                                                                                                                                                                                                                                                                                                                                                                                                                                                                                                                                                                                                
    g_1,g_2,g_3,\ldots
\]
will play the role of hidden microscopic variables.
For every $n\geq0$, define the binary variable
\begin{equation}                                                                                                                                                                                                                                                                                                                                                                                                                                                                                                                                                                                                                                
    X_n(g)=g_n.
\end{equation}
As discussed in Section~\ref{sec:formal}, each such variable induces a
partition of the microscopic state space:
\begin{equation}                                                                                                                                                                                                                                                                                                                                                                                                                                                                                                                                                                                                                                
    P_n
    =
    \left\{
    \{g\in\Omega:g_n=0\},
    \{g\in\Omega:g_n=1\}
    \right\}.
    \label{eq:Pn}
\end{equation}
The partition $P_n$ therefore retains exactly one microscopic distinction:
whether the $n$th coordinate is equal to $0$ or $1$.
We take
\begin{equation}                                                                                                                                                                                                                                                                                                                                                                                                                                                                                                                                                                                                                                
    \mathfrak P
    =
    \{P_n\}_{n\geq0}
    \label{eq:coordinate-vocabulary}
\end{equation}
as our candidate macroscopic vocabulary.
This vocabulary is manifestly effective. Given an integer $n$, the
corresponding partition is specified simply by asking for the value of the
$n$th binary coordinate. No search through the space of partitions is
required, and every member of the vocabulary has a finite description.

The observable we wish to predict is fixed once and for all:
\begin{equation}                                                                                                                                                                                                                                                                                                                                                                                                                                                                                                                                                                                                                                
    h(g)=g_0.
    \label{eq:observable}
\end{equation}
Thus the prediction problem concerns only the value of the readout coordinate.
The remaining ingredient is a mechanism by which different hidden
coordinates can become relevant to that prediction.

We introduce a context space $\mathcal C$ together with a selector
\begin{equation}                                                                                                                                                                                                                                                                                                                                                                                                                                                                                                                                                                                                                                
    \sigma:
    \mathcal C
    \longrightarrow
    \N_{\geq1}\cup\{\bot\}.
    \label{eq:sigma}
\end{equation}
The symbol $\bot$ means that no hidden coordinate is selected.
The role of the context is simple: if                                                                                                                                                                                                                                                                                                                                                                                                                                        
$\sigma(c)=n$,
then the next value of the readout coordinate will be copied from $g_n$. If                                                                                                                                                                                                                                                                                                                                                                                                                                            
$\sigma(c)=\bot$,
the readout is set to zero.
Define
\begin{equation}                                                                                                                                                                                                                                                                                                                                                                                                                                                                                                                                                                                                                                
    \rho(c,g)
    =
    \begin{cases}                                                                                                                                                                                                                                                                                                                                                                                                                                                                                                                                                                                                                                
        g_{\sigma(c)},
         & 
        \sigma(c)\neq\bot,
        \\[1mm]
        0,
         & 
        \sigma(c)=\bot.
    \end{cases}
    \label{eq:rho}
\end{equation}
The microscopic update rule is then
\begin{equation}                                                                                                                                                                                                                                                                                                                                                                                                                                                                                                                                                                                                                                
    G(c,g)_0
    =
    \rho(c,g),
    \qquad
    G(c,g)_n
    =
    g_n
    \quad
    (n\geq1).
    \label{eq:G-model}
\end{equation}
The hidden coordinates therefore never change. The only effect of the
context is to determine which hidden coordinate, if any, is copied into the
observable register at the next step.

This simple construction is useful because it makes the relation between
microscopic variables and macroscopic relevance completely transparent. If a
future context selects $g_n$, then knowledge of $g_n$ is necessary for
predicting the subsequent observable value. If no relevant context selects
$g_n$, then that coordinate can be ignored for the prediction task.

Consider, for example, a single context $c$ satisfying                                                                                                                                                                                                                                                                                                                                                                                                                                          
$\sigma(c)=3$.
Starting from
\[                                                                                                                                                                                                                                                                                                                                                                                                                                                                                                                                                                                                                                
    g=(g_0,g_1,g_2,g_3,\ldots),
\]
one application of the microscopic dynamics gives
\[                                                                                                                                                                                                                                                                                                                                                                                                                                                                                                                                                                                                                                
    G(c,g)
    =
    (g_3,g_1,g_2,g_3,\ldots).
\]
Since the next observable value is $g_3$, two microscopic states that differ
only in their third hidden coordinate can no longer be regarded as
macroscopically equivalent for this prediction problem. The partition $P_3$
must be retained.

The same reasoning extends to an arbitrary finite prediction horizon.
Let
\begin{equation}                                                                                                                                                                                                                                                                                                                                                                                                                                                                                                                                                                                                                                
    \mathbf c
    =
    (c_0,c_1,\ldots,c_{H-1})
\end{equation}
be a context word of length $H$. Since the hidden coordinates remain fixed
under the dynamics, the observable sequence generated from an initial state
$g$ is
\begin{equation}                                                                                                                                                                                                                                                                                                                                                                                                                                                                                                                                                                                                                                
    \Psi_{H,\mathbf c}(g)
    =
    \bigl(
    g_0,
    \rho(c_0,g),
    \rho(c_1,g),
    \ldots,
    \rho(c_{H-1},g)
    \bigr).
    \label{eq:model-prediction-map}
\end{equation}
The only hidden coordinates on which this sequence can depend are those
selected somewhere in the context word. Define
\begin{equation}                                                                                                                                                                                                                                                                                                                                                                                                                                                                                                                                                                                                                                
    S_H(\mathbf c)
    =
    \left\{
    \sigma(c_k):
    0\leq k<H,\;
    \sigma(c_k)\neq\bot
    \right\}.
    \label{eq:SH-word}
\end{equation}
The set $S_H(\mathbf c)$ contains the indices of all hidden coordinates that
can influence the observable during the chosen prediction horizon.
Two microscopic states $g$ and $g'$ therefore give identical predictions if
and only if they agree on the initial readout coordinate and on every hidden
coordinate belonging to $S_H(\mathbf c)$:
\begin{equation}                                                                                                                                                                                                                                                                                                                                                                                                                                                                                                                                                                                                                                
    g\sim_{H,\mathbf c}g'
\end{equation}
if and only if
\begin{equation}                                                                                                                                                                                                                                                                                                                                                                                                                                                                                                                                                                                                                                
    g_0=g_0'
    \qquad\text{and}\qquad
    g_n=g_n'
    \quad
    \text{for every }
    n\in S_H(\mathbf c).
    \label{eq:model-equivalence}
\end{equation}
The predictive partition is consequently
\begin{equation}                                                                                                                                                                                                                                                                                                                                                                                                                                                                                                                                                                                                                                
    \Pi_H(\mathbf c)
    =
    \CommonRef
    \left(
    \{P_0\}
    \cup
    \{P_n:n\in S_H(\mathbf c)\}
    \right).
    \label{eq:model-Pi}
\end{equation}
Equation~\eqref{eq:model-Pi} is the central observation of the construction. It says that
the minimal macroscopic description needed for prediction consists simply of
the readout variable together with the finitely many hidden coordinates that
the context word makes relevant.

For example, suppose that $H=3$ and the successive contexts select
\[                                                                                                                                                                                                                                                                                                                                                                                                                                                                                                                                                                                                                                
    \sigma(c_0)=2,
    \qquad
    \sigma(c_1)=5,
    \qquad
    \sigma(c_2)=2.
\]
Then
\[                                                                                                                                                                                                                                                                                                                                                                                                                                                                                                                                                                                                                                
    S_3(\mathbf c)=\{2,5\},
\]
and
\[                                                                                                                                                                                                                                                                                                                                                                                                                                                                                                                                                                                                                              
    \Pi_3(\mathbf c)
    =
    \CommonRef(\{P_0,P_2,P_5\}).
\]
Although the coordinate $g_2$ is consulted twice, it represents only one
macroscopic distinction. The prediction problem therefore requires knowledge
of precisely three variables:                                                                                                                                                                                                                                                                                                                                                                                                                                      
$g_0, g_2, g_5$.

Notice also why only finitely many members of the vocabulary appear in each
predictive description. The prediction horizon $H$ is finite, and at most one
coordinate can be selected at each step. Consequently,
\begin{equation}                                                                                                                                                                                                                                                                                                                                                                                                                                                                                                                                                                                                                                
    |S_H(\mathbf c)|\leq H.
\end{equation}
Every predictive partition is therefore a finite common refinement of
coordinate partitions.

We can now establish the formal prestatability of the entire predictive
architecture.

\begin{proposition}[Formal prestatability of the predictive class]
    \label{prop:predictive-formal-prestatability}
    For every finite prediction horizon $H$, the coordinate vocabulary
    \[                                                                                                                                                                                                                                                                                                                                                                                                                                                                                                                                                                                                                                
        \mathfrak P
        =
        \{P_n\}_{n\geq0}
    \]
    is effective and complete for the predictive admissible class
    \[                                                                                                                                                                                                                                                                                                                                                                                                                                                                                                                                                                                                                                
        \mathcal A_H
        =
        \left\{
        \Pi_H(\mathbf c):
        \mathbf c\in\mathcal C^H
        \right\}.
    \]
\end{proposition}

\begin{proof}                                                                                                                                                                                                                                                                                                                                                                                                                                                                                                                                                                                                                                
    The vocabulary is effective because each partition $P_n$ is specified by
    the finite index $n$.
    For any context word $\mathbf c$ of length $H$, the set
    $S_H(\mathbf c)$ contains at most $H$ indices. Equation
    \eqref{eq:model-Pi} therefore expresses $\Pi_H(\mathbf c)$ as a common
    refinement of finitely many members of $\mathfrak P$. Hence
    \[                                                                                                                                                                                                                                                                                                                                                                                                                                                                                                                                                                                                                                
        \Pi_H(\mathbf c)
        \in
        \langle\mathfrak P\rangle_{\mathrm{fin}}
    \]
    for every $\mathbf c\in\mathcal C^H$, and therefore
    \[                                                                                                                                                                                                                                                                                                                                                                                                                                                                                                                                                                                                                                
        \mathcal A_H
        \subseteq
        \langle\mathfrak P\rangle_{\mathrm{fin}}.
    \]
    The vocabulary is thus complete for $\mathcal A_H$.
\end{proof}

The significance of Proposition~\ref{prop:predictive-formal-prestatability} should be emphasized. The microscopic
space $\Omega$ is infinite, and the full partition space                                                                                                                                                                                                                                                                                                                                                                                                                                                           
$\Part(\Omega)$
is not exhaustively formally pre-statable, as shown in
Proposition~\ref{prop:failure-exhaustive-formal}.
But the prediction problem does not require access to arbitrary partitions of
$\Omega$. Its admissible macroscopic descriptions form a much smaller class,
and every member of that class can be constructed from the explicit
coordinate vocabulary
\[                                                                                                                                                                                                                                                                                                                                                                                                                                                                                                                                                                                                                                
    P_0,P_1,P_2,\ldots.
\]
Thus the predictive model suffers from no formal obstruction. Every
elementary distinction that could ever be needed by the prediction task has
already been specified.
What remains undetermined is which of those distinctions the future dynamics
will actually require.

To formulate this more explicitly, suppose that a context evolution                                                                                                                                                                                                                                                                                        
$F:\mathcal C\longrightarrow\mathcal C$
and an initial context $c_0$ are given. At time $t$, the relevant context word
is
\begin{equation}                                                                                                                                                                                                                                                                                                                                                                                                                                                                                                                                                                                                                               
    (c_t,c_{t+1},\ldots,c_{t+H-1}),
    \qquad
    c_t=F^t(c_0),
\end{equation}
and the corresponding set of selected coordinates is
\begin{equation}                                                                                                                                                                                                                                                                                                                                                                                                                                                                                                                                                                                                                                
    S_{H,t}^{F}
    =
    S_H(c_t,c_{t+1},\ldots,c_{t+H-1}).
    \label{eq:SH-time}
\end{equation}
For every $n\geq1$, the variable $P_n$ is relevant at time $t$ precisely when                                                                             
$n\in S_{H,t}^{F}$.
The future relevance set therefore takes the particularly simple form
\begin{equation}                                                                                                                                                                                                                                                                                                                                                                                                                                                                                                                                                                                                                                
    \mathcal R_H^{F}(c_0)
    =
    \left\{
    n\geq1:
    n\in S_{H,t}^{F}
    \text{ for some }t\geq0
    \right\}.
    \label{eq:R-model}
\end{equation}
The readout partition $P_0$ is always present and will be omitted from this
set, since its relevance is trivial.
The problem of dynamical prestatability has thus been reduced to a concrete
question: given an index $n$, can we determine whether the future context
trajectory will ever select the hidden coordinate $g_n$?

The next section shows that the answer depends entirely on the structure
of the context dynamics. We will compare two systems built on exactly the
same microscopic state space, observable, candidate vocabulary, and update
rule. In the first, the complete future relevance set is immediately
decidable. In the second, determining whether a candidate variable will ever
become relevant is equivalent to the halting problem.

\section{Two Context Dynamics}
\label{sec:dynamics}

We are now in a position to ask the central dynamical question. The predictive
architecture of the previous section is fixed. The microscopic state space is                                                                                                                                                                                                                                                                                                                                                                                   
$\Omega=\F_2^{\N}$,
the observable is always                                                                                                                                                                                                                                                                                                                                                                                  
$h(g)=g_0$,
and the candidate macroscopic vocabulary is
\[                                                                                                                                                                                                                                                                                                                                                                                                                                                                                                                                                
    \mathfrak P=\{P_0,P_1,P_2,\ldots\}.
\]
A context can make the variable $P_n$ relevant only by selecting the hidden
coordinate $g_n$ and copying its value into the observable register. Every
possible variable that could be needed by the prediction problem has
therefore already been specified.
What remains is to determine which coordinates the future context trajectory
will actually select.

We compare two context dynamics. In the first, the sequence of selected
coordinates can be written down explicitly in advance. In the second, the
selection of coordinates is governed by a universal computation.

Consider first the simplest possible case. Let the context trajectory be                                                                                                                                                                                                                                                                                                                                                                                    
$c_0^{\mathrm R},c_1^{\mathrm R},c_2^{\mathrm R},\ldots,$
with
\begin{equation}                                                                                                                                                                                                                                                                                                                                                                                                                                                                                                                                                
    F_{\mathrm R}(c_t^{\mathrm R})=c_{t+1}^{\mathrm R},
    \label{eq:FR}
\end{equation}
and define the selector by
\begin{equation}                                                                                                                                                                                                                                                                                                                                                                                                                                                                                                                                                
    \sigma(c_t^{\mathrm R})=t+1.
    \label{eq:sigma-R}
\end{equation}
Thus the successive contexts select
\[                                                                                                                                                                                                                                                                                                                                                                                                                                                                                                                                                
    g_1,g_2,g_3,g_4,\ldots
\]
in order.
Thus, at time $t$, prediction over a horizon
$H$ depends on the coordinates
\begin{equation*}                                                                                                                                                                                                                                                                                                                                                                                                                                                                                                                                                
    g_{t+1},g_{t+2},\ldots,g_{t+H}.
\end{equation*}
Together with the initial observable $g_0$, the prediction map is therefore
\begin{equation}                                                                                                                                                                                                                                                                                                                                                                                                                                                                                                                                                
    \Psi^{\mathrm R}_{H,t}(g)
    =
    \bigl(
    g_0,
    g_{t+1},
    g_{t+2},
    \ldots,
    g_{t+H}
    \bigr).
    \label{eq:Psi-R}
\end{equation}
The corresponding macroscopic description is
\begin{equation}                                                                                                                                                                                                                                                                                                                                                                                                                                                                                                                                                
    \Pi^{\mathrm R}_{H,t}
    =
    \CommonRef
    \left(
    \{P_0,P_{t+1},P_{t+2},\ldots,P_{t+H}\}
    \right).
    \label{eq:Pi-R}
\end{equation}
For example, if $H=2$, the relevant variables evolve as
\[                                                                                                                                                                                                                                                                                                                                                                                                                                                                                                                                                
    (P_0,P_1,P_2)
    \longrightarrow
    (P_0,P_2,P_3)
    \longrightarrow
    (P_0,P_3,P_4)
    \longrightarrow\cdots.
\]
The macroscopic description therefore changes continually. Variables that
were relevant cease to be needed, while previously irrelevant variables enter
the description.
Yet there is no difficulty in stating the future pattern. Every hidden
coordinate is eventually selected. Hence, for every $H\geq1$,
\begin{equation}                                                                                                                                                                                                                                                                                                                                                                                                                                                                                                                                                
    \mathcal R_H^{\mathrm R}
    =
    \N_{\geq1}.
    \label{eq:R-R}
\end{equation}
Given any candidate variable $P_n$, we can therefore answer immediately that
it will eventually become relevant. The system is dynamically pre-statable.

We now change only the process that determines which coordinates are
selected.
Let
\begin{equation}                                                                                                                                                                                                                                                                                                                                                                                                                                                                                                                                                
    Q_1,Q_2,Q_3,\ldots
\end{equation}
be an effective enumeration of programs for a universal model of computation.
For each $n$, consider the computation obtained by running $Q_n$ on its own
index:                               
$Q_n(n)$.
Define the diagonal halting set
\begin{equation}                                                                                                                                                                                                                                                                                                                                                                                                                                                                                                                                                
    K
    =
    \left\{
    n\geq1:
    Q_n(n)\text{ eventually halts}
    \right\}.
    \label{eq:K}
\end{equation}
Membership in $K$ is undecidable \cite{Turing1936}. 

We now let the context dynamics simulate all of these computations in
parallel by means of a standard dovetailing procedure. The idea of
dovetailing is simply to ensure that no individual computation is neglected.
One elementary step is performed for one program, then for another, and so
on, repeatedly returning to each computation. If some $Q_n(n)$ eventually
halts, the dovetailer will therefore discover this after a finite number of
steps.
Let                                 
$c_0^{\mathrm U}$
denote the initial state of the dovetailing computation, and let                            
$\operatorname{Next}(c^{\mathrm U})$
denote one elementary dovetailing step. The context dynamics is
\begin{equation}                                                                                                                                                                                                                                                                                                                                                                                                                                                                                                                                                
    F_{\mathrm U}(c^{\mathrm U})
    =
    \operatorname{Next}(c^{\mathrm U}).
    \label{eq:FU}
\end{equation}
The context state contains the finite information required to continue the
dovetailing computation. Hence both $F_{\mathrm U}$ and each individual
finite step of the context evolution are effectively computable.

We now connect this computation to the predictive architecture of the
previous section. Whenever a dovetailing step discovers for the first time
that                                           
$Q_n(n)$
has halted, the corresponding context selects the coordinate $g_n$. At all
other steps, no hidden coordinate is selected. Thus
\begin{equation}                                                                                                                                                                                                                                                                                                                                                                                                                                                                                                                                                
    \sigma(c^{\mathrm U})
    =
    \begin{cases}                                                                                                                                                                                                                                                                                                                                                                                                                                                                                                                                                
        n,
         & 
        \text{if the current step first discovers that }Q_n(n)\text{ halts},
        \\[1mm]
        \bot,
         & 
        \text{otherwise}.
    \end{cases}
    \label{eq:sigma-U}
\end{equation}
Recall that $\sigma(c^{\mathrm U})=\bot$ simply means that no hidden coordinate becomes
relevant at that step. In such a context the readout is assigned the fixed
value specified in Section~\ref{sec:model}.

Now consider a particular candidate variable $P_n$.
Suppose first that $Q_n(n)$ eventually halts. Because the dovetailing schedule
is fair, this halt will eventually be discovered. At that moment,                                                  
$\sigma(c_t^{\mathrm U})=n$
for some finite time $t$. The microscopic update then copies                                                 
$g_n$
into the observable register. Predicting the subsequent observable value
therefore requires knowledge of $g_n$, and the macroscopic description must
retain $P_n$. Hence
\[                                                                                                                                                                                                                                                                                                                                                                                                                                                                                                                                                
    Q_n(n)\text{ halts}
    \quad\Longrightarrow\quad
    P_n\text{ eventually becomes relevant}.
\]
Conversely, suppose that $Q_n(n)$ never halts. The dovetailing computation can
continue forever, but it will never discover a halt for this program.
Consequently,
\[                                                                                                                                                                                                                                                                                                                                                                                                                                                                                                                                                
    \sigma(c_t^{\mathrm U})\neq n
\]
for every $t$. The observable never depends on $g_n$, and the prediction
problem never requires the distinction $P_n$. Therefore
\[                                                                                                                                                                                                                                                                                                                                                                                                                                                                                                                                                
    Q_n(n)\text{ does not halt}
    \quad\Longrightarrow\quad
    P_n\text{ never becomes relevant}.
\]
Combining the two directions gives
\begin{equation}                                                                                                                                                                                                                                                                                                                                                                                                                                                                                                                                                
    P_n\text{ becomes relevant at some future time}
    \quad\Longleftrightarrow\quad
    Q_n(n)\text{ halts}.
    \label{eq:halting-equivalence}
\end{equation}
Equivalently, for every finite prediction horizon $H\geq1$,
\begin{equation}                                                                                                                                                                                                                                                                                                                                                                                                                                                                                                                                                
    \mathcal R_H^{\mathrm U}
    =
    K.
    \label{eq:R-U}
\end{equation}
The future relevance problem is therefore undecidable.

\begin{proposition}[Dynamical unprestatability]
    \label{prop:dynamical-unprestatability}
    For every fixed finite prediction horizon $H\geq1$, the universal model
    is formally pre-statable but dynamically unprestatable. Its candidate
    macroscopic vocabulary is effective and complete, but there exists no
    algorithm that decides, for every $n$, whether $P_n$ will ever become
    relevant.
\end{proposition}

\begin{proof}                                                                                                                                                                                                                                                                                                                                                                                                                                                                                                                                                
    Formal prestatability was established in
    Proposition~\ref{prop:predictive-formal-prestatability}: every
    admissible predictive partition is a finite refinement of members of
    \[                                                                                                                                                                                                                                                                                                                                                                                                                                                                                                                                                
        \mathfrak P=\{P_n\}_{n\geq0}.
    \]
    Suppose now that there existed a terminating algorithm which, given $n$,
    decided whether $P_n$ would ever become relevant. By
    Eq.~\eqref{eq:halting-equivalence}, the same algorithm would decide whether
    $Q_n(n)$ eventually halts. This would decide membership in the diagonal
    halting set $K$, contradicting the undecidability of the halting problem.
\end{proof}

The contrast between the two models is particularly instructive. In both
cases the microscopic phase space is completely specified. The observable is
completely specified. The elementary macroscopic vocabulary is completely
specified. Every admissible macroscopic description is constructed from that
vocabulary, and the evolution at every specified finite time can be computed.
The difference lies solely in the global structure of the context dynamics.

In the reducible model, the sequence of selected coordinates is
\[                                                                                                                                                                                                                                                                                                                                                                                                                                                                                                                                                
    1,2,3,4,\ldots,
\]
so the complete future relevance set can be stated immediately.
In the universal model, a coordinate $n$ is selected precisely when a
corresponding computation is discovered to halt. Every positive instance can
eventually be witnessed: if $P_n$ becomes relevant, running the dynamics long
enough will eventually reveal that fact. What cannot be determined in
general is that $P_n$ will \emph{never} become relevant, since this would
require knowing that the corresponding computation will never halt.

\section{Conclusion}
\label{sec:conclusions}

At first sight, macroscopic unprestatability appears paradoxical. If the microscopic phase space is complete and the laws governing its evolution are fully specified, how can anything about the system's future macroscopic description remain unknown or even unprestatable? After all, every macroscopic property must ultimately be grounded in the microscopic states, and no new microscopic possibilities can emerge outside the original phase space. One might therefore expect microscopic completeness to guarantee macroscopic prestatability.

The resolution lies in recognizing that microscopic states and macroscopic descriptions are fundamentally different objects. Macroscopic descriptions are ways of organizing and distinguishing microscopic states, and their space of possibilities can be vastly larger than the microscopic phase space itself. Indeed, even a finite phase space admits exponentially many subsets and an even larger number of possible partitions. For infinite spaces, the difference is no longer merely quantitative, but qualitative: the two spaces have different cardinality. Thus, knowing the complete microscopic phase space does not imply having a practically or algorithmically accessible catalogue of its possible macroscopic descriptions. The latter may be formally specified without being effectively enumerable.

Our construction takes this distinction further by showing that even when the relevant macroscopic descriptions are formally pre-statable, determining which of them will become relevant during the system's evolution can be undecidable. Formal prestatability therefore does not guarantee dynamical prestatability, namely that the future macroscopic organization can be anticipated.

\end{document}